\documentclass{emulateapj}
\submitted{{\it in November 2008 issue of PASP, 120, 1161 }}

\shortauthors{West \& Hawley}
\shorttitle{$\chi$ Values}
\begin{document}

\title{$\chi$ Values for Blue Emission Lines in M Dwarfs}

\author{Andrew A. West}
\affil{Astronomy Department, University of California, Berkeley, CA}
\affil{MIT Kavli Institute for Astrophysics and Space Research, Cambridge, MA}
\email{aaw@space.mit.edu}

\author{Suzanne L. Hawley}
\affil{Department of Astronomy, University of Washington, Seattle, WA}
\email{slh@astro.washington.edu}

\begin{abstract} 
  We compute $\chi$ values for blue emission lines in active
  M dwarfs.  Using flux-calibrated spectra from nearby M dwarfs and
  spectral M dwarf templates from SDSS, we derive analytic relations
  that describe how the $\chi$ values for the CaII H and K as well as
  the H$\beta$, H$\gamma$, H$\delta$, H$\epsilon$ and H8 Balmer
  emission lines vary as a function of spectral type and color.  The
  $\chi$ values are useful for numerous M dwarf studies where the
  intrinsic luminosity of emission lines cannot be estimated due to
  uncertain distances and/or non-flux-calibrated spectra.  We use
  these results to estimate the mean properties of blue emission lines
  in active field M dwarfs from SDSS.
\end{abstract}


\section{Introduction}
The domination of M dwarfs in the Galaxy's stellar census makes them
ideal tracers of the kinematics, structure, and evolution of the Milky
Way. Many of these stars are also host to large magnetic fields that
act to heat the upper atmosphere and give rise to magnetic activity
(as often traced by chromospheric emission lines).  Observations of
magnetic activity put important constraints on the internal structure,
the relationship between magnetic field generation and rotation,
atmospheric models, and the ages of M dwarfs.  Although low-mass stars
are intrinsically faint, recent large surveys include an unprecedented
number of M dwarfs. The Sloan Digital Sky Survey \citep[SDSS;][]{DR6}
in particular has identified $\sim$30 million M dwarfs in the
photometric database (Bochanski et al., in prep.) and over 40,000
M dwarfs with spectra \citep{W08}.  \citet{W04,W06} and \citet{Boo07}
used these spectra to statistically examine the magnetic activity
properties (as traced by H$\alpha$) of field M dwarfs.  Because the
SDSS sample traces a large range of physical and dynamic properties,
it provides an important laboratory for studying how magnetic activity
changes as a function of mass, metallicity, rotation, and stellar age
\citep{W08}.

Magnetic activity is often quantified by computing the ratio of the
luminosity in an emission line (typically H$\alpha$) to the bolometric
luminosity of the star \citep[L$_{\rm{H}\alpha}$/L$_{bol}$;][]{Reid95,Hawley96,Burgasser02,W04}.
This quantity allows for comparison of stars of different temperatures,
where changes in the continua can severely affect the derived
equivalent widths.  Although the activity-bolometric ratio is strictly
distance independent, the challenges associated with obtaining
accurate bolometric fluxes (namely good bolometric corrections for
individual stars) have resulted in studies that use a color or
spectral type dependent bolometric luminosity
\citep[e.g.][]{Hawley96}, and thus still require a distance to solve
for the line emission luminosity.  For H$\alpha$, the line luminosity
is often calculated using the equivalent width of the emission line,
and the luminosity of the continuum near H$\alpha$; the latter can be
measured directly from a flux-calibrated spectrum and a measured
stellar distance.  Alternatively, the continuum luminosity can be
estimated from the photometric colors \citep{Reid95}. Because of the
difficulty in calculating individual bolometric corrections for large
samples of stars, the fact that many spectra are not flux-calibrated
(e.g. high resolution echelle data and data taken under
non-photometric conditions) and that distance estimates to most M
dwarfs are still a major source of systematic uncertainty
\citep{Covey08}, an alternative approach to calculating the
activity-bolometric ratio is required.

\citet[hereafter WHW04]{WHW04} and later \citet{W05}, introduced a
novel method for calculating L$_{\rm{H}\alpha}$/L$_{bol}$ by removing
many of the intermediate steps and requiring only a measurement of the
equivalent width of the emission line (EW) and either the color or the
spectral type of the star.   WHW04\nocite{WHW04}, defined a
sample of nearby stars with known distances (from trigonometric
parallax), derived bolometric corrections and used flux-calibrated spectra
to solve for the ratio of the H$\alpha$ continuum region luminosity to
the bolometric luminosity as a function of both spectral type and
color.  This ratio is dubbed $\chi$,

\begin{equation}
\chi = L_{\rm{H}\alpha}(continuum)/L_{bol},
\end{equation}

\noindent and when multiplied by the EW of the
H$\alpha$ emission line yields the following L$_{\rm{H}\alpha}$/L$_{bol}$ ratio:

\begin{equation}
L_{\rm{H}\alpha}/L_{bol}=\chi \times EW_{\rm{H}\alpha} ~(\rm{\AA}).
\end{equation}

\noindent The power of the $\chi$ method
(WHW04\nocite{WHW04}) is that it does not require a flux-calibrated
spectrum to obtain L$_{\rm{H}\alpha}$/L$_{bol}$, and it is
distance independent.  Although previous studies had applied similar
strategies for obtaining L$_{\rm{H}\alpha}$/L$_{bol}$
\citep{Reid95,Hawley96,MB03}, they did not explicitly report recipes
for obtaining ``$\chi$'' and they relied on colors and photometric or
spectroscopic parallax relations, or in the latter case, spectral models,
to obtain L$_{\rm{H}\alpha}$(continuum).
WHW04 was the first study to use flux-calibrated
spectra to determine L$_{\rm{H}\alpha}$(continuum) and to present 
empirical $\chi$ values in H$\alpha$ for the entire M dwarf sequence.

Recently, \citet{Reiners08} derived $\chi$ values for H$\alpha$ from model spectra
as a function of effective temperature. Although the \citet{Reiners08}
$\chi$ relations are similar to those in WHW04 for stars hotter than 2600 K
($\sim$M6), they do not produce the plateau seen in the late-type
empirically derived $\chi$s from WHW04.  This discrepancy is likely due
to problems with the spectral models of late-type dwarfs, which do not
reproduce the correct optical colors and magnitudes of
M dwarfs \citep{Covey08}.  The difference in derived
log(L$_{\rm{H}\alpha}$/L$_{bol}$) values using the two methods rises to
0.6 dex at a spectral type of M9.  We prefer to rely on the measured
data rather than models, and adopt the WHW04 $\chi$ values for H$\alpha$
in our analysis.

The WHW04\nocite{WHW04} $\chi$s have shown great utility in
computing L$_{\rm{H}\alpha}$/L$_{bol}$ in numerous studies
\citep[e.g.,][]{W04,Silvestri06,Boo07,Reiners08}, but H$\alpha$ is not the
only emission line produced by an active chromosphere. In fact,
H$\alpha$ may not always be the best tracer of activity as compared to
other lines \citep{Browning08, Walkowicz08}.  Traditionally, there
have been several limitations in using the bluer emission lines,
namely that the signal-to-noise ratio
(S/N) in the blue part of an M dwarf spectrum is low because the
continuum emission peaks in the near-infrared, and CCDs have been more sensitive in
the red.  Several recent studies have examined magnetic activity as traced
by the Ca II H and K emission lines and higher order
Balmer transitions including H$\beta$, H$\gamma$, H$\delta$,
H$\epsilon$ and H8 \citep{Rauscher06, Boo07,Browning08, Walkowicz08},
but star-to-star comparisons have been limited by the lack of line
specific $\chi$ conversions to activity-bolometric luminosity ratios.

In this paper, we expand upon the WHW04\nocite{WHW04} study, using
similar methods to compute $\chi$ values for the Ca II
H and K emission lines as well as the H$\beta$, H$\gamma$, H$\delta$,
H$\epsilon$ and H8 Balmer emission lines.  We describe the data and
our techniques in \S2. In \S3 we derive $\chi$ for each emission line
as a function of both color and spectral type and use these values to
investigate the mean emission line properties of active M dwarfs.  We
discuss the results in \S4.

\section{Data and Analysis}
We used two complementary techniques and datasets to derive a single
set of $\chi$ values for the blue emission lines in M dwarfs.  The
first technique uses flux-calibrated blue spectra for a sample of
nearby stars with good parallaxes and distances. 
The second made use of the
\citet{Boo07} SDSS spectral templates, which provide high S/N average
spectra at each M spectral type, flux-calibrated on a relative scale.
These two methods were combined to derive $\chi$ values as a function
of both color and spectral type and are described in detail below.

\begin{deluxetable*}{lccccccccc}
\tablewidth{0pt}
\tablewidth{0pt}
\tablecolumns{10} 
\tabletypesize{\scriptsize}
\tablecaption{Nearby M Dwarfs with Blue Spectra}
\renewcommand{\arraystretch}{.6}
\tablehead{
\colhead{Name}&
\colhead{Sp. Type}&
\colhead{parallax (mas)}&
\colhead{$B$\tablenotemark{a}}&
\colhead{$V$\tablenotemark{a}}&
\colhead{$R$\tablenotemark{a}}&
\colhead{$I$\tablenotemark{a}}&
\colhead{$J$\tablenotemark{b}}&
\colhead{$H$\tablenotemark{b}}&
\colhead{$K_S$\tablenotemark{b}}}
\startdata
Gl 277b  & M3.5  &  87.15 &  13.30 & 11.78 & 10.62 & 9.07 & 7.57 &6.99 & 6.76\\
Gl 277a  & M2.5 &     87.61 &  12.03 & 10.57 & 9.52 & 8.15 & 6.77 &6.18 & 5.93\\
CU Cnc  & M3.5 &  78.05 &12.83 &12.05 &11.4 &\nodata &7.51 &6.90 &6.60\\
BD+33 1646b & M3 & 48.26 & 13.2 & 11.4 & \nodata & \nodata & 8.00 & 7.36&7.16\\
Gl 473AB & M5/M7 &227.0 & 14.30 & 12.46 & 10.90 & 8.92 & 6.99 & 6.40 & 6.04\\
AD Leo & M3 &213.0 & 10.85 & 9.32 & 8.23 & 6.81 & 5.45 & 4.84 & 4.59\\ 
DT Vir & M0.5 & 87.50 & 11.23 & 9.76 & 8.81 & 7.71 & 6.44 & 5.79 & 5.58\\
Gl 490A & M0.5& 55.27 & 12.2 & 10.5 & 9.73 & 8.8 & 7.40 & 6.73 & 6.55\\ 
Gl 725B & M3.5 &284.48 &  11.47 & 9.69 & 8.56 & 7.13 & 5.72 & 5.20 & 5.00\\
CE Boo & M2.5  &101.91 & 11.68 & 10.2 & 9.4 & 8.5 & 6.63 & 5.99 & 5.77\\
Gl 644AB & M3/M4 & 174.22 & 10.60 & 9.02 & 7.92 & 6.55 & 5.27 & 4.78 & 4.40\\
Gl 725A & M3 & 280.28 & 10.44 & 8.90 & 7.83 & 6.44 & 5.19 & 4.74 & 4.43\\ 
Gl 182 & M0.5 &  37.50 &  11.48 & 10.07 & 9.18 & 8.24 & 7.12 & 6.45 & 6.26\\
Gl 234AB & M4.5/M8 & 242.88 &  12.80 & 11.08 & 9.77 & 8.06 &   8.10 & 7.47 & 7.21\\
Gl 268AB & M4.5/M6 & 157.23 &  13.19 & 11.49 & 10.16 & 8.44 & 6.73 & 6.15 &5.85\\
YZ CMi & M4.5 & 168.59 & 12.76 & 11.15 & 9.89 & 8.2 & 6.58 & 6.01 & 5.70\\
CW UMa & M3.5 &68.3 &  14.02 & 12.38 & 11.2 & 9.7 & 8.30 & 7.76 & 7.50\\
Gl 685 & M0.5 &70.95 &  11.45 & 9.97 & 9.2 & 8.3 & 6.88 & 6.27 & 6.07\\
\enddata 
\tablecomments{M dwarfs from the spectroscopic sample of \citet{PH89}}
\tablenotetext{a}{Optical photometry from \citet{Bessell91}, \citet{Koen02},
  \citet{Leggett92} and SIMBAD}
\tablenotetext{b}{IR photometry from 2MASS \citep{2MASS}}
\label{tab:stars}
\end{deluxetable*}

\subsection{Nearby Stars with Flux-Calibrated Spectra}
We used the blue spectra of 18 nearby M dwarfs from \citet[hereafter
PH89]{PH89}.  The observations were made using the Cassegrain
spectrograph on the 2.1m Struve telescope at McDonald Observatory, 
with resolution R$\sim$1,000 and spectral range 3600-4500 \AA.  
Our analysis used the
PH89\nocite{PH89} spectra as well as a few additional unpublished
spectra from the same program.  Because most of the PH89\nocite{PH89}
stars were observed at multiple epochs, we chose the highest S/N
spectrum for each star to use in our analysis.  A list of the stars
and their properties can be found in Table \ref{tab:stars}.  The optical
and infrared photometry listed in Table \ref{tab:stars} were compiled from
\citet{Bessell91}, \citet{Koen02}, \citet{Leggett92},
\citet[2MASS]{2MASS} and
SIMBAD\footnote{SIMBAD can be accessed online at http://simbad.u-strasbg.fr/Simbad.}.  The
parallaxes were taken from the Hipparcos catalog \citep{HIPPARCOS}.
Close binary systems (Gl 473AB, Gl 644AB, Gl 234AB, Gl268AB) were
not separated. Their combined spectra are ascribed to the earlier type
star in the system.

\begin{deluxetable}{lcccccc}
\tablewidth{0pt}
\tablecolumns{7} 
\tabletypesize{\scriptsize}
\tablecaption{Continuum Regions}
\renewcommand{\arraystretch}{.6}
\tablehead{
\colhead{Emission Line}&
\colhead{Region 1 (\AA)}&
\colhead{Region 2 (\AA)}}
\startdata
H$\alpha$ & 6555-6560 & 6570-6575\\
H$\beta$ &4840-4850 & 4875-4885\\
H$\gamma$ &4310-4330 & 4350-4370\\
H$\delta$ &4075-4095 & 4110-4130\\
H8 &3865-3885 & 3895-3915\\
CaII K/CaII H/H$\epsilon$ &3974-3976 & 3952.7-3956\\
\enddata
\label{tab:cont}
\end{deluxetable}

Bolometric corrections for each star were computed using the $K$-band
relations from \citet{Leggett96,Leggett01} as a function of $I-K$
colors. In two cases $I_C$ magnitudes were not available and the $J-K$
relations were used.  All of the bolometric corrections were derived
after first transforming the $K_S$ 2MASS magnitudes to $K_{UKIRT}$
magnitudes using the \citet{Carpenter01} relations.  Bolometric
luminosities were obtained assuming $M_{\rm{bol},\odot}=4.64$ \citep{Schmidt82}.

Continuum fluxes were measured by taking the mean continuum flux near
emission lines in each spectrum.  Table \ref{tab:cont} gives the
wavelength range for each continuum region.  The CaII H, CaII K and
H$\epsilon$ lines have the same continuum region
\citep[as in][]{Rauscher06}. The continuum fluxes were converted
to luminosity using the distance to each star.  By dividing the
continuum luminosity by the bolometric luminosity, we calculated the
$\chi$ value for each emission line region for each star.
We caution that to properly use the $\chi$ values presented in this paper,
equivalent widths should be computed using the continuum regions
given in Table \ref{tab:cont}.  All equivalent widths presented
in this paper have been computed with these continuum regions, and
users should do the same with their data.

\begin{figure*}[h]
\plotone{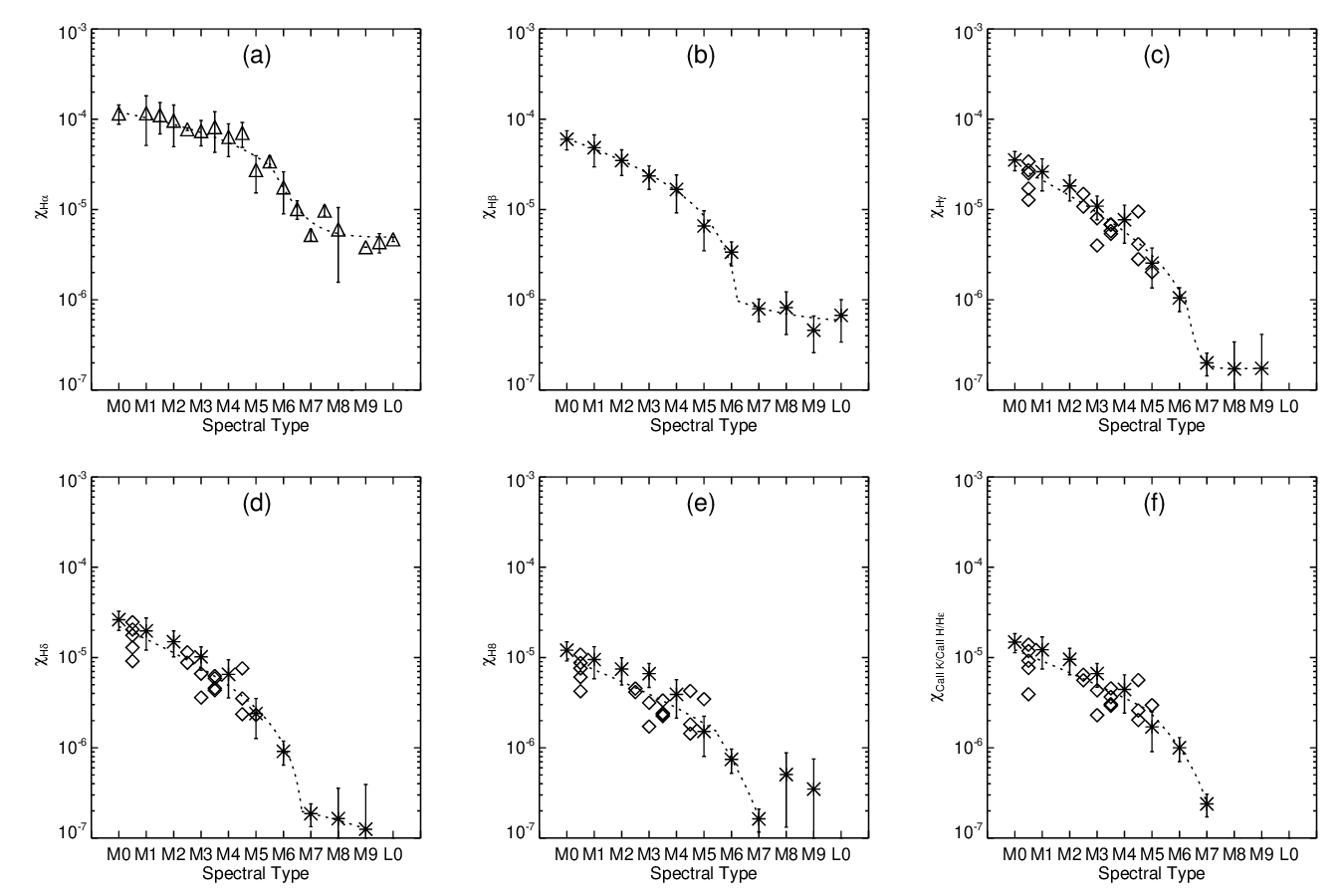}
\caption{$\chi$ as a function of spectral type for the (a) H$\alpha$,
  (b) H$\beta$, (c) H$\gamma$, (d) H$\delta$, (e) H8, and (f) H$\epsilon$,
  CaII H and CaII K emission lines. The $\chi$ values were
  derived using results from the study of WHW04 (triangles; H$\alpha$), the
  \citet{Boo07} SDSS spectral templates (asterisks), and the flux-calibrated
  spectra of nearby stars (PH89; diamonds).  An empirical relation
  (dotted lines) for
  each $\chi$ value can be found using the coefficients in Table
  \ref{tab:spt}.}
\label{fig:spt}
\end{figure*}

\begin{deluxetable*}{lcccccc}[h]
\tablewidth{0pt}
\tablecolumns{7} 
\tabletypesize{\scriptsize}
\tablecaption{ $\chi$ Values}
\renewcommand{\arraystretch}{.6}
\tablehead{
\colhead{Sp. Type}&
\colhead{$\chi_{\rm{H}\alpha}$ ($\times 10^{-4}$)\tablenotemark{a}}&
\colhead{$\chi_{\rm{H}\beta}$ ($\times 10^{-4}$)}&
\colhead{$\chi_{\rm{H}\gamma}$ ($\times 10^{-4}$)}&
\colhead{$\chi_{\rm{H}\delta}$ ($\times 10^{-4}$)}&
\colhead{$\chi_{\rm{H8}}$ ($\times 10^{-4}$)}&
\colhead{$\chi_{\rm{CaII K/CaII H/H}\epsilon}$ ($\times 10^{-4}$)\tablenotemark{b}}}
\startdata
  M0   &   1.160 (0.277)  &  0.600 (0.143)  & 0.253 (0.090)  & 0.185 (0.067)  &  0.082 (0.029)  &   0.102 (0.041)\\
    M1 &    1.160 (0.451)  & 0.484 (0.187)  & 0.238 (0.077)  & 0.174 (0.056) &     0.078 (0.024)  &  0.098 (0.036)\\
    M2  &   0.966 (0.306) &  0.349 (0.111)  &  0.146 (0.038)  & 0.117 (0.031) &   0.054 (0.018)   &  0.072 (0.021)\\
    M3  &   0.738 (0.216)  &   0.235 (0.069) &  0.082 (0.034) &  0.069 (0.027)  &  0.033 (0.015)  &  0.044 (0.016)\\
    M4  &   0.637 (0.286)  &   0.167 (0.075)  &  0.061 (0.021)  & 0.051  (0.017)  &  0.027 (0.010)  &  0.036 (0.012)\\
    M5   &   0.274 (0.128) &  0.066 (0.031)  & 0.042 (0.031)  &  0.036 (0.023)  &  0.025 (0.013)  &    0.030 (0.016)\\
    M6  &   0.176 (0.052)  &  0.034 (0.010)  &  0.011 (0.003) &  0.009 (0.003)   & 0.007 (0.002) &  0.010 (0.003)\\
    M7  &  0.052 (0.015) &   0.008 (0.002) &  0.002 (0.001) &  0.002 (0.001) & 0.002 (0.001)  & 0.002 (0.001)\\
    M8  &  0.060 (0.029) &  0.008 (0.004) &  0.002 (0.002) &  0.002 (0.002) &  0.005 (0.004) & 0.001 (0.002)\\
    M9 &   0.038 (0.011) &  0.005 (0.002) & 0.002 (0.001) &  0.001 (0.001) &   0.003 (0.004) &  \nodata  (\nodata)\\
   L0  &  0.047 (0.007)  & 0.007 (0.003) & 0.001 (0.001) & \nodata (\nodata) & \nodata  (\nodata) & \nodata  (\nodata)\\
\enddata
\tablecomments{Uncertainties derived from the spread in $\chi$ are
  given in parentheses.}
\tablenotetext{a}{H$\alpha$ $\chi$ values computed from WHW04\nocite{WHW04}.}
\tablenotetext{b}{The $\chi$ values for CaII K, CaII H and
  H$\epsilon$ are the same because the continuum regions used to compute the
  equivalent widths are the same for all three lines.}
\label{tab:chi}
\end{deluxetable*}

\subsection{Bochanski Templates}
Because the PH89\nocite{PH89} spectra only include M dwarfs with
spectral types M0-M5, we supplemented our analysis with the SDSS
spectral templates of \citet{Boo07}.  \citet{Boo07} used over 4000
SDSS spectra to construct mean templates of M0-L0 dwarfs that have
both higher S/N and higher resolution (R$\sim$10,000) than their
native SDSS components.  The higher resolution was achieved by using a
spectral drizzle technique that involves co-adding a large number of spectra
after first adjusting each spectrum to zero radial velocity on a sub-pixel
grid (for more details see \citealp{Boo07}).  Although the template spectra are
not on an absolute flux scale, they are on a normalized relative flux
scale that is accurate to better than 4\% \citep{Boo07}.  The templates therefore
provide high S/N measurements of the continuum flux (on a relative
scale) for each of the M dwarf spectral types.  The templates
are further divided into categories of H$\alpha$ activity.  We
examined the active and inactive templates at each spectral type and
found no continuum variation in the active versus inactive spectra.
We therefore chose to use the combined (active plus inactive) templates
for our $\chi$ analysis.

We measured the continuum region around the H$\alpha$ emission line at
each spectral type (see Table \ref{tab:cont}) and used the $\chi$
values from WHW04\nocite{WHW04} to calibrate each template spectrum to
units of L/L$_{bol}$.  We measured the continuum regions around each
emission line for all of the templates in the same manner as the
PH89\nocite{PH89} spectra.  Because of the WHW04\nocite{WHW04}
calibration, the mean continuum values are direct measurements of
$\chi$.  Uncertainties in the $\chi$ values were computed from the
spread of the composite spectra used to create the \citet{Boo07}
templates.

\begin{figure*}[h]
\plotone{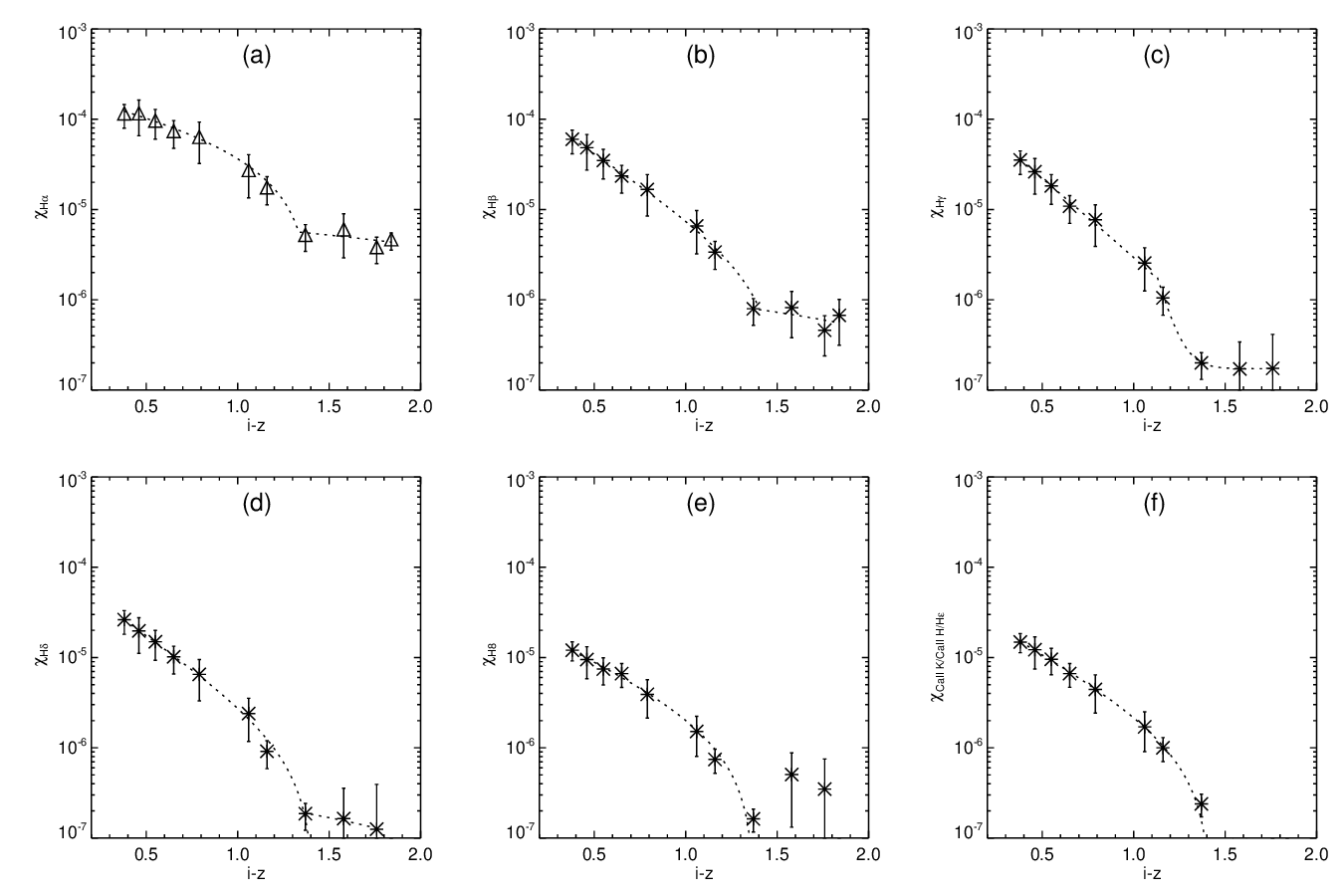}
\caption{$\chi$ as a function of $i-z$ color for the (a) H$\alpha$,
 (b) H$\beta$, (c) H$\gamma$, (d) H$\delta$, (e) H8, and (f) H$\epsilon$,
  CaII H and CaII K emission lines. The $\chi$ values were
  derived from the study of WHW04 (triangles; H$\alpha$) and the SDSS
  spectral templates of \cite[asterisks]{Boo07}.  An empirical
  relation (dotted lines) for each
  $\chi$ value can be found using the coefficients in Table
  \ref{tab:i_z}.}
\label{fig:i_z} 
\end{figure*}

\begin{figure*}[h]
\plotone{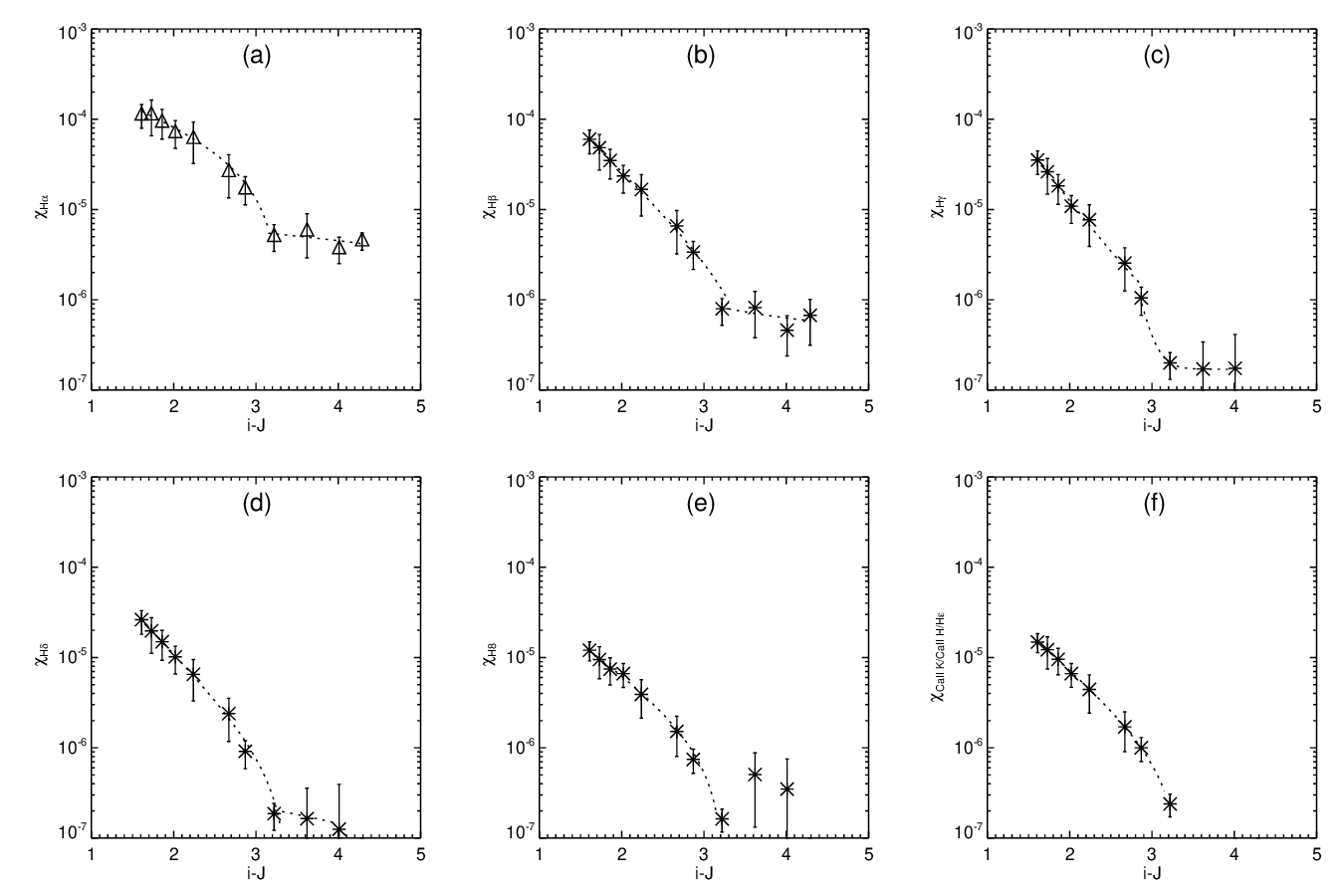}
\caption{$\chi$ as a function of $i-J$ color for the (a) H$\alpha$,
 (b) H$\beta$, (c) H$\gamma$, (d) H$\delta$, (e) H8, and (f) H$\epsilon$,
  CaII H and CaII K emission lines. The $\chi$ values were
  derived from the study of WHW04 (triangles; H$\alpha$) and the
  spectral templates of \citet[asterisks]{Boo07}.  An empirical
  relation (dotted lines) for each
  $\chi$ value can be found using the coefficients in Table
  \ref{tab:i_J}.}
\label{fig:i_j} 
\end{figure*}

\begin{figure*}[h]
\plotone{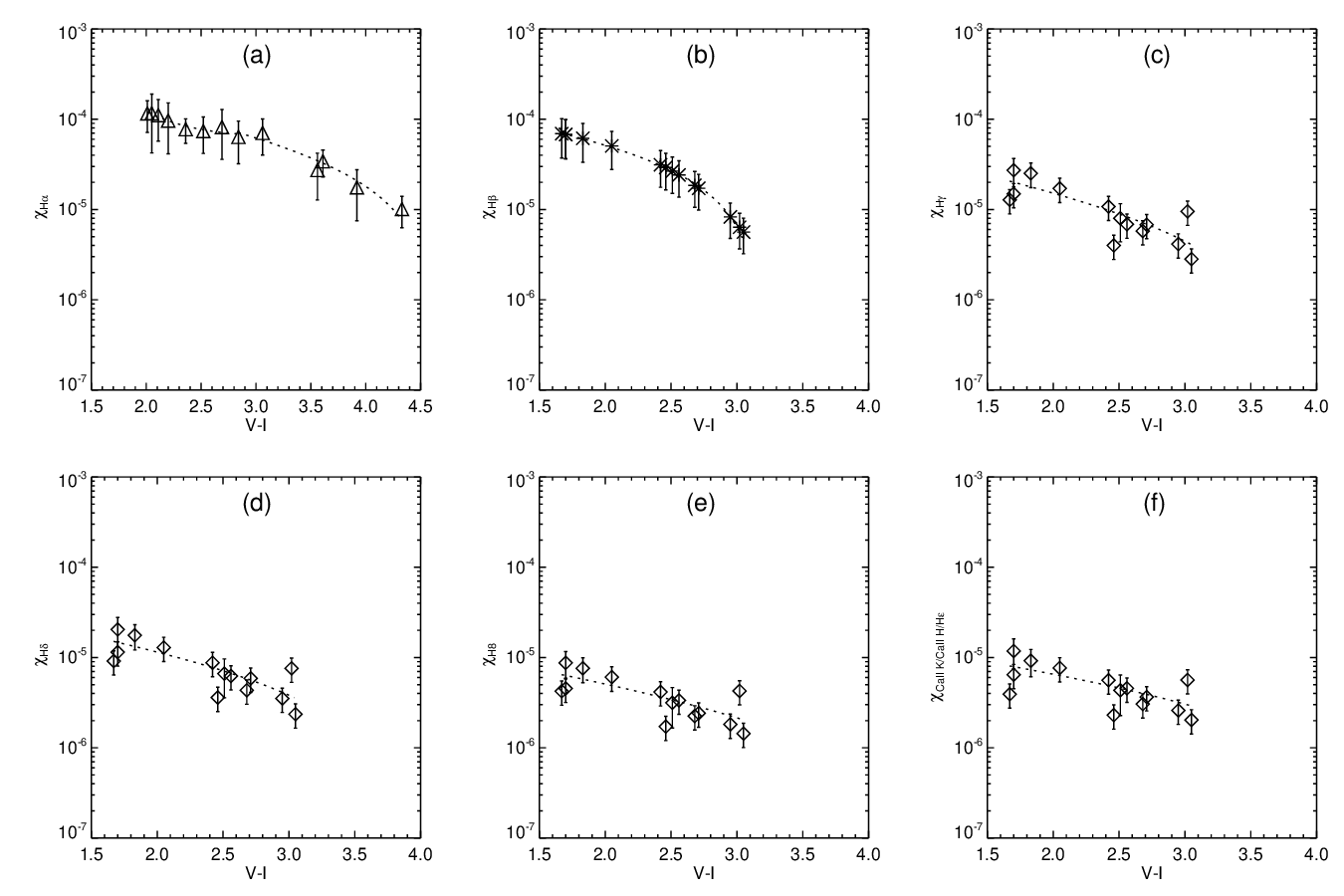}
\caption{$\chi$ as a function of $V-I$ color for the (a) H$\alpha$,
 (b) H$\beta$, (c) H$\gamma$, (d) H$\delta$, (e) H8, and (f) H$\epsilon$,
  CaII H and CaII K emission lines. The $\chi$ values were
  derived from the study of WHW04 (triangles; H$\alpha$), the spectral
  templates of \citet[asterisks, H$\beta$]{Boo07}, and the flux-calibrated spectra of
  nearby stars (PH89; diamonds).  An empirical relation (dotted lines)
  for each $\chi$ value
  can be found using the coefficients in Table \ref{tab:V_I}.}
\label{fig:V_I} 
\end{figure*}

\begin{deluxetable*}{lccccc}[h]
\tablewidth{0pt}
\tablecolumns{6} 
\tabletypesize{\scriptsize}
\tablecaption{$\chi$ vs. Spectral Type}
\renewcommand{\arraystretch}{.6}
\tablehead{
\colhead{$\chi$}&
\colhead{a}&
\colhead{b}&
\colhead{c}&
\colhead{Range}&
\colhead{$\sigma$$\chi$}}
\startdata
$\chi_{\rm{H}\alpha}$ & 0.54186 & 3.33$\times10^4$ & -0.54173 & M0-M5.5
(0-5.5) & 3.31$\times10^{-5}$\\
$\chi_{\rm{H}\alpha}$ & 0.193 & 0.621 & 4.92$\times10^{-6}$ & M5.5-L0
(5.5-10) & 4.55$\times10^{-6}$\\
$\chi_{\rm{H}\beta}$ & 9.49$\times10^{-5}$ & 6.21 &
-3.39$\times10^{-5}$ & M0-M6.5 (0-6.5) & 3.17$\times10^{-6}$\\
$\chi_{\rm{H}\beta}$ & 1.37$\times10^{-5}$ & 1.80 & 5.39$\times10^{-7}$ & M6.5-L0
(6.5-10) & 2.25$\times10^{-7}$\\
$\chi_{\rm{H}\gamma}$ & 3.42$\times10^{-5}$ & 3.04 &
-3.50$\times10^{-6}$ & M0-M6.5 (0-6.5) & 2.14$\times10^{-5}$\\
$\chi_{\rm{H}\gamma}$ & 1.64$\times10^{6}$ & 0.22 & 1.73$\times10^{-7}$ & M6.5-M9
(6.5-9) & 2.00$\times10^{-7}$\\
$\chi_{\rm{H}\delta}$ & 2.65$\times10^{-5}$ & 3.84 &
-4.47$\times10^{-6}$ & M0-M6.5 (0-6.5) & 1.63$\times10^{-5}$\\
$\chi_{\rm{H}\delta}$ & 4.63$\times10^{-4}$ & 1.52$\times10^{4}$ & -4.619$\times10^{-4}$ & M6.5-M9
(6.5-9) & 2.00$\times10^{-7}$\\
$\chi_{\rm{H}8}$ & 1.11$\times10^{-5}$ & 4.01 & -1.329$\times10^{-6}$ &
M0-M5.5 (0-5.5) & 8.31$\times10^{-6}$\\
$\chi_{\rm{H}8}$ & 6.448$\times10^{-4}$ & 0.91 & -1.26$\times10^{-7}$ & M5.5-M7
(5.5-7) & 1.38$\times10^{-6}$\\
$\chi_{\rm{CaII K/CaII H/H}\epsilon}$ & 1.334$\times10^{-5}$ & 3.82 &
-1.124$\times10^{-6}$ & M0-M5 (0-5) & 1.07$\times10^{-5}$\\
$\chi_{\rm{CaII K/CaII H/H}\epsilon}$ & 5.233$\times10^{-5}$ & 1.77 & -7.629$\times10^{-7}$ & M5-M7
(5-7) & 8.95$\times10^{-7}$\\
\enddata
\tablecomments{Coefficients for calculating $\chi$ as a function of
  spectral type using the equation $\chi =  a e^{-SpT/b}+c$. Spectral
  types are specified by integer values from M0=0 to L0=10.}
\label{tab:spt}
\end{deluxetable*}

\begin{deluxetable*}{lccccc}[h]
\tablewidth{0pt}
\tablecolumns{6} 
\tabletypesize{\scriptsize}
\tablecaption{$\chi$ vs. $i-z$}
\renewcommand{\arraystretch}{.6}
\tablehead{
\colhead{$\chi$}&
\colhead{a}&
\colhead{b}&
\colhead{c}&
\colhead{Range}&
\colhead{$\sigma$$\chi$}}
\startdata
$\chi_{\rm{H}\alpha}$ & 2.92$\times10^{-4}$ & 1.19  &
-8.96$\times10^{-5}$ & 0.38-1.37 & 1.31$\times10^{-5}$\\
$\chi_{\rm{H}\alpha}$ & 0.011788 & 4.66$\times10^3$ &
-1.1779$\times10^{-2}$ & 1.37-1.84 & 1.32$\times10^{-6}$\\
$\chi_{\rm{H}\beta}$ & 2.02$\times10^{-4}$ & 0.324 &
-1.86$\times10^{-6}$ & 0.38-1.37 & 2.68$\times10^{-6}$\\
$\chi_{\rm{H}\beta}$ & 2.378$\times10^{-4}$ & 4.61$\times10^2$ &
-2.363$\times10^{-4}$ & 1.37-1.84 & 2.15$\times10^{-7}$\\
$\chi_{\rm{H}\gamma}$ & 1.62$\times10^{-4}$ & 0.252 &
-1.38$\times10^{-7}$ & 0.38-1.10 & 1.61$\times10^{-6}$\\
$\chi_{\rm{H}\gamma}$ & 1.76$\times10^{2}$ & 0.061 &
1.72$\times10^{-7}$ & 1.10-1.76 & 2.00$\times10^{-7}$\\
$\chi_{\rm{H}\delta}$ & 9.17$\times10^{-5}$ & 0.311 &
-9.58$\times10^{-7}$ & 0.38-1.37 & 6.68$\times10^{-7}$\\
$\chi_{\rm{H}\delta}$ & 2.366$\times10^{-4}$ & 1.53$\times10^{3}$ &
-2.362$\times10^{-4}$ & 1.37-1.76 & 2.00$\times10^{-7}$\\
$\chi_{\rm{H}8}$ & 3.02$\times10^{-5}$ & 0.473 & -1.64$\times10^{-6}$
& 0.38-1.37 & 8.38$\times10^{-7}$\\
$\chi_{\rm{CaII K/CaII H/H}\epsilon}$ & 4.31$\times10^{-5}$ & 0.385 &
-1.07$\times10^{-6}$ & 0.38-1.37 & 4.87$\times10^{-7}$\\
\enddata
\tablecomments{Coefficients for calculating $\chi$ as a function of
  $i-z$ color using the equation $\chi =  a e^{-(i-z)/b}+c$.}
\label{tab:i_z}
\end{deluxetable*}

\begin{deluxetable*}{lccccc}[h]
\tablewidth{0pt}
\tablecolumns{6} 
\tabletypesize{\scriptsize}
\tablecaption{$\chi$ vs. $i-J$}
\renewcommand{\arraystretch}{.6}
\tablehead{
\colhead{$\chi$}&
\colhead{a}&
\colhead{b}&
\colhead{c}&
\colhead{Range}&
\colhead{$\sigma\chi$}}
\startdata
$\chi_{\rm{H}\alpha}$ & 5.25$\times10^{-4}$ & 1.53 &
-6.09$\times10^{-5}$ & 1.61-3.22 & 1.29$\times10^{-5}$\\
$\chi_{\rm{H}\alpha}$ & 2.257$\times10^{-3}$ & 2.01$\times10^3$ &
-2.248$\times10^{-3}$ & 3.22-4.29 & 1.33$\times10^{-6}$\\
$\chi_{\rm{H}\beta}$ & 1.79$\times10^{-3}$ & 0.477 &
-8.18$\times10^{-7}$ & 1.61-3.22 & 2.75$\times10^{-6}$\\
$\chi_{\rm{H}\beta}$ & 6.46$\times10^{-6}$ & 1.14 &
4.36$\times10^{-7}$ & 3.22-4.29 & 2.21$\times10^{-7}$\\
$\chi_{\rm{H}\gamma}$ & 2.72$\times10^{-3}$ & 0.370 &
3.03$\times10^{-7}$ & 1.61-2.85 & 1.58$\times10^{-6}$\\
$\chi_{\rm{H}\gamma}$ & 2.71$\times10^{5}$ & 0.108 &
1.70$\times10^{-7}$ & 2.85-4.01 & 2.00$\times10^{-7}$\\
$\chi_{\rm{H}\delta}$ & 8.83$\times10^{-4}$ & 0.460  &
-5.47$\times10^{-7}$& 1.61-3.22 & 6.06$\times10^{-7}$\\
$\chi_{\rm{H}\delta}$ & 1.33085$\times10^{-3}$ & 1.73$\times10^4$ &
-1.3304$\times10^{-3}$ & 3.22-4.01 & 2.00$\times10^{-7}$\\
$\chi_{\rm{H}8}$ & 1.37$\times10^{-4}$ & 0.685 & -1.19$\times10^{-6}$
& 1.61-3.22 & 8.61$\times10^{-7}$\\
$\chi_{\rm{CaII K/CaII H/H}\epsilon}$ & 2.74$\times10^{-4}$ & 0.563 &
-6.85$\times10^{-7}$ & 1.61-3.22 & 4.10$\times10^{-7}$\\
\enddata
\tablecomments{Coefficients for calculating $\chi$ as a function of
  $i-J$ color using the equation $\chi =  a e^{-(i-J)/b}+c$.}
\label{tab:i_J}
\end{deluxetable*}

\begin{deluxetable*}{lccccc}[h]
\tablewidth{0pt}
\tablecolumns{6} 
\tabletypesize{\scriptsize}
\tablecaption{$\chi$ vs. $V-I$}
\renewcommand{\arraystretch}{.6}
\tablehead{
\colhead{$\chi$}&
\colhead{a}&
\colhead{b}&
\colhead{c}&
\colhead{Range}&
\colhead{$\sigma\chi$}}
\startdata
$\chi_{\rm{H}\alpha}$ & 1.92$\times10^{-2}$ & 0.345 &
6.36$\times10^{-5}$ & 2.01-2.84 & 1.54$\times10^{-5}$\\
$\chi_{\rm{H}\alpha}$ & 4.87$\times10^{-4}$ & 2.07 &
-5.25$\times10^{-5}$ & 2.84-4.33 & 1.63$\times10^{-5}$\\
$\chi_{\rm{H}\beta}$ & 3.12$\times10^{-4}$ & 3.15 &
-1.14$\times10^{-4}$ & 1.67-3.05 & 2.71$\times10^{-6}$\\
$\chi_{\rm{H}\gamma}$ & 8.38$\times10^{-5}$ & 1.49 &
-6.74$\times10^{-6}$ & 1.67-3.05 & 1.62$\times10^{-5}$\\
$\chi_{\rm{H}\delta}$ & 5.42$\times10^{-5}$ & 1.83  &
-6.70$\times10^{-6}$& 1.67-3.05 & 1.17$\times10^{-5}$\\
$\chi_{\rm{H}8}$ & 2.08$\times10^{-5}$ & 1.83 & -1.89$\times10^{-6}$ &
1.67-3.05 & 5.30$\times10^{-6}$\\
$\chi_{\rm{CaII K/CaII H/H}\epsilon}$ & 2.36$\times10^{-5}$ & 2.26 &
-3.21$\times10^{-6}$ & 1.67-3.05 & 7.45$\times10^{-6}$\\
\enddata
\tablecomments{Coefficients for calculating $\chi$ as a function of
  $V-I$ color using the equation $\chi =  a e^{-(V-I)/b}+c$.}
\label{tab:V_I}
\end{deluxetable*}

\begin{figure*}[h]
\plotone{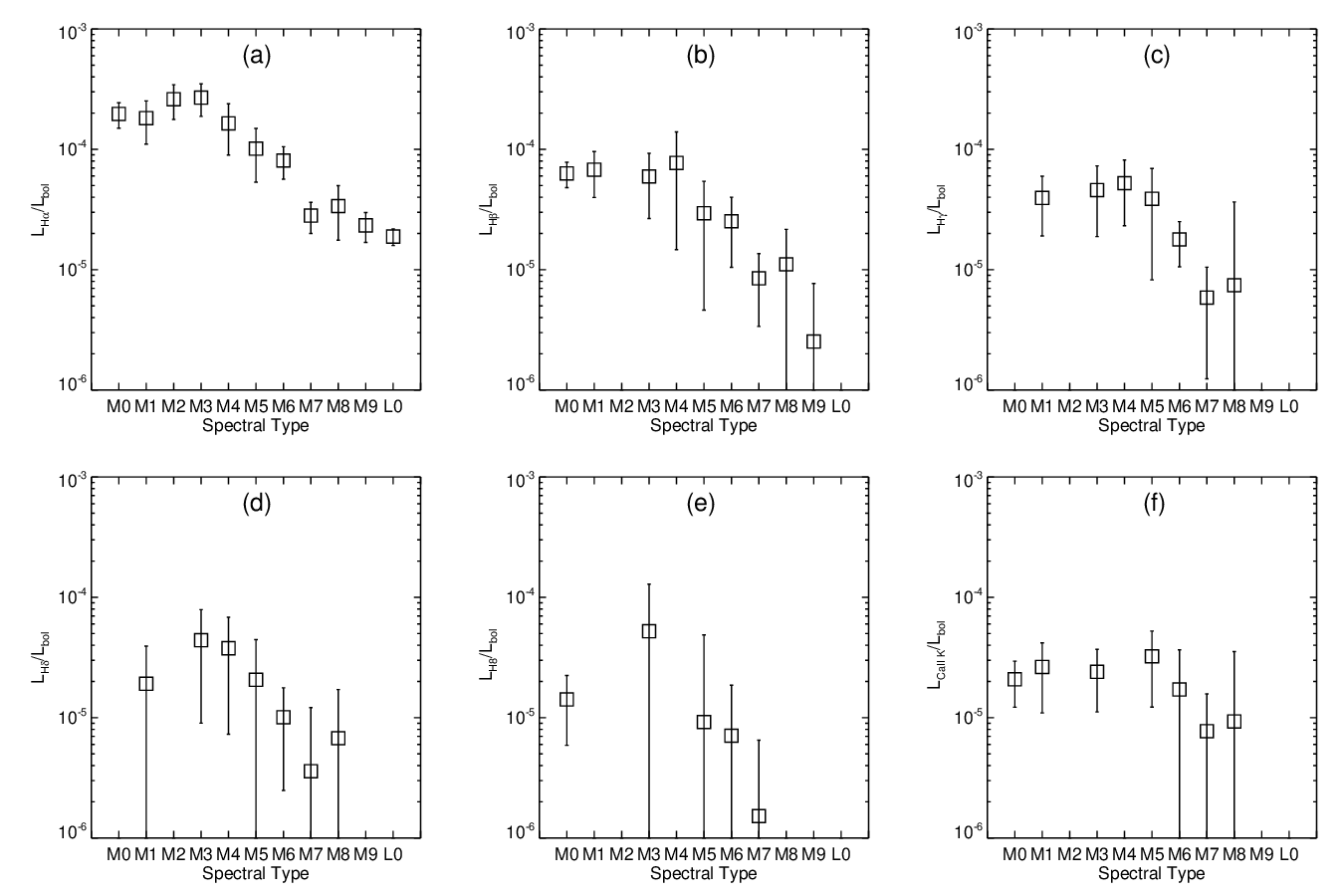}
\caption{L$_{line}$/L$_{bol}$ as a function of spectral type for the
 (a) H$\alpha$, (b) H$\beta$, (c) H$\gamma$, (d) H$\delta$, (e) H8,
  and (f) CaII K emission lines. These values were computed using
  equivalent width measurements from the median composite active SDSS
  template spectra of \citet{Boo07} and the $\chi$ values presented in
  this paper.  The relations
  show that the fraction of the bolometric luminosity that is produced
  in a given emission line increases monotonically with wavelength at
  all spectral types. Also, all relations are consistent with the
  H$\alpha$ study of \citet{W04} that showed a constant
  L$_{\rm{H}\alpha}$/L$_{bol}$ value from M0-M4 with a decrease
  occurring at a spectral type $\sim$M5.  In addition, the L$_{\rm{CaII
    K}}$/L$_{bol}$ relation traces the L$_{\rm{CaII K}}$/L$_{bol}$ found for
  H$\alpha$ active M dwarfs in \citet{Browning08}. }
\label{fig:lbol} 
\end{figure*}

\section{Results}
The resulting $\chi$ values are shown as a function of spectral type
in Figure \ref{fig:spt}.  The H$\alpha$ values (upper-left; triangles)
are from the WHW04\nocite{WHW04} study.  Asterisks denote the $\chi$
values derived from the Bochanski templates, and diamonds show the
$\chi$ values computed from the nearby blue spectra.  There is
excellent agreement between the two different methods.  Table
\ref{tab:chi} gives the mean $\chi$ values as a function of spectral
type with uncertainties representing the spread of $\chi$ at each
spectral type (in parentheses).  Interestingly, Figure \ref{fig:spt} shows 
a plateau in $\chi$ values over the spectral type range M7-L0 in every
emission line for which we have data.  This
was previously seen, though not discussed, in WHW04\nocite{WHW04} for H$\alpha$.  
This striking feature is evidently produced
by a balance between the decreasing temperature (bolometric
luminosity) and decreasing optical continuum flux in these
late-type dwarfs.  We note again that these empirical results
disagree with the \citet{Reiners08} values derived using model
spectra, which we attribute to deficiencies in the models.  
The details of this plateau are left for future analysis.

Figures \ref{fig:i_z}-\ref{fig:V_I} show the $\chi$ values as a
function of $i-z$, $i-J$, and $V-I$ colors.  For the $i-z$ and $i-J$
relations, we transformed from spectral type to color using the mean
$i-z$ and $i-J$ colors for M dwarfs reported by \citet{W08}.  The
$V-I$ vs. $\chi$ relations for H$\gamma$, H$\delta$, H8, along with the
CaIIK, CaIIH, and H$\epsilon$, were obtained using the observed colors of the
individual nearby stars in PH89\nocite{PH89} (see Table \ref{tab:stars})
and the measured $\chi$
values discussed above, while the H$\alpha$ relation is taken from the
colors and $\chi$ values in WHW04\nocite{WHW04}.  The colors for the
H$\beta$ $V-I$ versus $\chi$ relation were derived by transforming the mean
$i-z$ colors for each spectral type to $V-I$ using the $r-i$ vs. $V-I$
transformation of \citet{Davenport06} and the following linear
transformation between $r-i$ and $i-z$ that we found from the
\citet{W08} sample: 

\begin{equation}
i-z=0.54(r-i)+0.01.
\end{equation}

\noindent The $V-I$ vs. $\chi$ relations do not extend beyond $V-I$ $>$
3.05 ($\sim$M5; except for H$\alpha$) due to the lack of late-type M
dwarfs with measured $V-I$ colors in the PH89\nocite{PH89} sample. 

We fit analytic relations to our derived $\chi$ values as a
function of spectral type, $i-z$, $i-J$ and $V-I$ color.  We adopt the
exponential form of WHW04:

\begin{equation}
\chi =  a e^{-V/b}+c,\\
\end{equation}

\noindent where $a$, $b$ and $c$ are derived coefficients and $V$ is
the associated variable (e.g. $i-z$, $i-J$).  All of the fits
were performed using a Levenberg-Marquardt least-squares method.
Tables \ref{tab:spt}-\ref{tab:V_I} give the resulting fit
coefficients for each $\chi$ value.  Most of the fits require 2
piecewise (but continuous) components; the tables specify the range
over which each fit component is valid.  In addition, the tables provide a
typical uncertainty in the derived $\chi$ for each fit component.

Using our derived $\chi$ values and the active templates of
\citet{Boo07}, we next computed L$_{line}$/L$_{bol}$ values as a function
of spectral type for the H$\alpha$, H$\beta$, H$\gamma$, H8 and CaII K
emission lines.  The \citet{Boo07} active templates consist of the
mean spectrum of up to several hundred active SDSS M dwarfs at each
spectral type.  We measured the EWs of the H$\alpha$, H$\beta$,
H$\gamma$, H8 and CaII K emission lines in each of the \citet{Boo07}
active templates, taking care to use the continuum regions from
Table \ref{tab:cont} for our measurements (see Section 2.1).  
Lines which were not in emission in a given template were omitted.
Combining the measured EWs with
our new $\chi$ values gives L$_{line}$/L$_{bol}$ values at
each spectral type for each line. 
These represent the average values of active field M dwarfs
measured by SDSS.   

Figure
\ref{fig:lbol} shows the new activity relations, which represent the mean
luminosity fractions of six different activity-tracing emission lines
for thousands of active field M dwarfs. 
The relations suggest that the
fraction of the bolometric luminosity that is produced in a given
emission line increases monotonically with wavelength at all spectral
types. In addition, the newly measured emission lines are consistent 
with the H$\alpha$ study of \citet{W04} that showed a constant value for
L$_{\rm{H}\alpha}$/L$_{bol}$ for early-mid M types
(M0-M4) with a decrease occurring near spectral type $\sim$M5.  
The L$_{\rm{CaII K}}$/L$_{bol}$ relation shown here agrees with
L$_{\rm{CaII K}}$/L$_{bol}$ results for the H$\alpha$ active 
M dwarfs from \citet{Browning08}.

\section{Discussion}
The $\chi$ values from WHW04\nocite{WHW04} provide a
distance-independent way to compare the H$\alpha$ activity of stars
with different spectral types.  We used a spectroscopic sample of
nearby M dwarfs from \citet{PH89} and the \citet{Boo07} templates to expand the
$\chi$ values to include the H$\beta$, H$\gamma$, H$\delta$,
H$\epsilon$, H8, CaII K, and CaII H emission lines.  We fit analytic
relations to all of the $\chi$ values (including H$\alpha$) as a
function of spectral type and color ($i-z$, $i-J$, and $V-I$).  The
new $\chi$ values are useful for analyzing both past and future M
dwarf data.  Magnetic activity can be easily quantified if studies include
EW measurements (measured using the continuum regions given
in Table \ref{tab:cont}) and either spectral types or colors.

We show that the percentage of bolometric luminosity emitted in the optical
emission lines of active-field M dwarfs increases
monotonically with increasing wavelength.  In addition, the emission
line luminosities relative to the bolometric luminosities are constant
in all measured lines for spectral types M0-M4, and then decrease at 
later spectral types, in agreement with the H$\alpha$ results in
\citet{W04}.  The L$_{\rm{CaII K}}$/L$_{bol}$ values are
consistent with those measured for the H$\alpha$ active stars from 
\citet{Browning08}.
Unfortunately, the H$\alpha$ inactive templates of \citet{Boo07} do
not have a sufficient S/N at each spectral type to systematically
compare the strength of the CaII K emission lines in active and
inactive stars.  However, a CaII K emission line is present in the M5
inactive template. The resulting mean L$_{\rm{CaII K}}$/L$_{bol}$ for the
inactive template is an order of magnitude lower than the (H$\alpha$)
active template.

Our $\chi$ values have already proven useful in the rotation-activity
analysis of \citet{Browning08} by allowing L$_{\rm{CaII K}}$/L$_{bol}$ and
L$_{\rm{CaII H}}$/L$_{bol}$ to be computed from echelle spectra. Future
improvements to our $\chi$ measurements can be made by creating a larger
spectroscopic sample of nearby M dwarfs with good parallaxes. 
The advent of synoptic telescopes like Pan-STARRS,
LSST and SIM will measure parallaxes for thousands of M dwarfs, many of
which will have SDSS (or other previously observed) spectra. These
samples will enable tighter constraints on $\chi$ and better inform
our understanding of magnetic activity in the majority of stellar
inhabitants of the Galaxy.

\acknowledgments The authors thank Matthew Browning,
Gibor Basri, Lucianne Walkowicz, Ansgar Reiners, and John Bochanski for
many fruitful conversations and suggestions that greatly improved the
quality of this manuscript.

\bibliographystyle{apj}


\end{document}